\documentclass{article}

\PassOptionsToPackage{numbers, compress}{natbib}

\usepackage[main, final]{arxiv_style}

\usepackage[utf8]{inputenc} 
\usepackage[T1]{fontenc}    
\usepackage{hyperref}       
\usepackage{url}            
\usepackage{booktabs}       
\usepackage{amsfonts}       
\usepackage{nicefrac}       
\usepackage{microtype}      
\usepackage{xcolor}         
\usepackage{amsmath, amssymb}
\usepackage{algorithm, algpseudocode}
\usepackage{bbm}             
\usepackage{graphicx}
\usepackage{multirow}
\usepackage[table]{xcolor}
\usepackage{wrapfig}
\usepackage{amsthm}
\usepackage{threeparttable}

\newtheorem{theorem}{Theorem}[section]

\title{Sequential Behavioral Watermarking for LLM Agents}

\author{%
  $^1$Hyeseon An, $^2$Shinwoo Park\thanks{This work was carried out while the author was at Yonsei University.}, $^1$Dongsu Kim, and $^1$Yo-Sub Han\thanks{Corresponding author} \\
  Department of Computer Science\\
  $^1$Yonsei University\\
  $^2$University of Luxembourg \\
  \texttt{\{\href{mailto:hsan@yonsei.ac.kr}{hsan},
  \href{mailto:123mki@yonsei.ac.kr}{123mki},
  \href{mailto:emmous@yonsei.ac.kr}{emmous}\}@yonsei.ac.kr} \\
  \href{mailto:shinwoo.park@uni.lu}{\texttt{shinwoo.park@uni.lu}}
}

\begin{document}

\maketitle

\begin{abstract}


LLM-based agents act through sequences of executable decisions, but their trajectories provide little evidence of which agent or policy produced them, making provenance, ownership, and unauthorized reuse difficult to establish from observed behavior alone. This motivates watermarking signals embedded directly into agent behavior rather than only into generated text, since text watermarking cannot capture the action-level decisions that define agent execution. Recent agent watermarking methods address this gap by moving the watermark from generated text to behavioral choices. However, by treating each action step as an independent trial, they overlook trajectory structure and become fragile when trajectories are perturbed, truncated, or observed without reliable alignment. We propose \texttt{SeqWM}, a sequential behavioral watermarking framework that embeds signals into history-conditioned transition patterns and verifies trajectories position-agnostically against random-key baselines. Experiments across diverse agent benchmarks and LLM backbones show that \texttt{SeqWM} consistently achieves reliable detection while preserving agent utility, and remains robust under trajectory corruption where round-indexed behavioral watermarks collapse.

\end{abstract}

\section{Introduction}
\label{sec:intro}

Large language models have moved beyond the role of text generators and now
serve as the decision-making core of autonomous agents that plan, invoke tools,
and act on external environments over long horizons. Systems such as Claude
Code~\citep{anthropic2025claudecode}, OpenAI's Deep Research~\citep{openai2025deepresearch}, and
Microsoft Copilot~\citep{microsoft2023copilot} are no longer experimental: they are deployed
into production workflows where the agent itself decides how to decompose tasks,
which tools to call, and in what order. This shift has two consequences that are
easy to state and increasingly hard to ignore. First, the locus of intellectual
property has moved from human-designed orchestration logic to the agent's
intrinsic decision-making capability, which is acquired through expensive
post-training on curated trajectory data~\citep{wang2026agentwm}. Second, the
trajectories themselves have become valuable assets: producing a single
high-quality SWE-Bench-style trajectory is reported to cost on the order of
\$100~\citep{pan2025swegym}, individual tool-use dialogues run roughly \$8
each~\citep{li2023apibank}, and curating the 120-task Mind2Web~2 benchmark required
over 1{,}000 hours of human labor~\citep{gou2025mind2web2}. Agents and the data that
train them have become economic objects whose unauthorized replication is now
both technically feasible and commercially attractive, while the infrastructure
for verifying their provenance has not kept up.

Across these deployments, a single technical problem recurs: given an observed
sequence of agent actions, can one verify which agent produced it? The same
primitive underlies a range of downstream concerns, from detecting impersonation
and coordinated manipulation on open social
platforms~\citep{huang2025agentguide,huang2026agentmark}, to attributing automated content to its
originating system, to defending proprietary agents from imitation attacks in
which an adversary harvests trajectories through ordinary API access and trains
a surrogate that recovers most of the victim's capability~\citep{wang2026agentwm}.
Trajectory datasets released for research face the parallel risk of being used
downstream without attribution~\citep{meng2026acthook}.


\paragraph{Why token-level watermarking does not transfer.}
The natural starting point is the rich literature on watermarking for
LLM-generated text, which biases token-level sampling toward a keyed subset of
the vocabulary~\citep{kirchenbauer2023watermark,dathathri2024scalable}. When applied to agents, however, this
approach is mismatched with the unit at which agents produce, and the unit at
which detectors observe, behavior. At generation time, a single high-level
decision such as which tool to call is realized through many tokens, and the
green/red bias on each individual token largely shapes syntactic surface form
rather than the decision itself; the watermark signal is diluted before it has
any meaningful effect on which action is taken. At verification time, the
situation is worse: a survey of 29 commercial agentic platforms reports that
82.8\% expose only the final action stream while concealing internal reasoning
traces and intermediate tokens~\citep{wang2026agentwm}, so whatever residual signal had
survived generation is largely discarded before reaching the detector.
Training-time alternatives that embed the signal directly into model weights
face their own obstacle, since most deployed agents are built on closed
pre-trained APIs that cannot be modified. The signal, in short, is lost
twice---once into the wrong substrate and once at the visibility
boundary---which suggests that watermarking for agents must operate at the level
of \emph{behavioral} decisions rather than tokens.

\paragraph{Existing behavioral watermarks.}
Recent work has begun to watermark LLM agents at the behavioral level, either
by biasing action choices, modifying trajectory data, or exploiting equivalent
tool-execution paths
\citep{huang2025agentguide,meng2026acthook,wang2026agentwm,huang2026agentmark}.
These approaches establish action-level provenance as a practical direction,
but they do not fully exploit the sequential structure of agent behavior. In
particular, the watermark evidence is still tied to individual decisions,
triggered actions, or fixed execution choices, rather than to the conditional
transition patterns that characterize how an agent acts over time.

\texttt{SeqWM} takes this sequential structure as the carrier of the watermark. Instead
of seeding the signal by absolute round index, we condition it on a window of
recent actions, making the watermark depend on local behavioral transitions.
This removes the need for strict positional alignment and makes deletion or
partial observation a local rather than global failure mode. We further add
multi-channel redundancy to strengthen the surviving evidence, and use
random-key calibration to obtain valid detection without assuming independent
sliding-window indicators. Together, these choices shift behavioral
watermarking from per-step action statistics to position-agnostic verification
of sequential behavior.

\paragraph{Contributions.}
We make the following contributions:
\begin{itemize}
    \item \textbf{Framing.} We recast agent behavioral watermarking as a sequential coding problem, and identify an asymmetry overlooked by prior work: while watermark generation is naturally chained through action history, verification need not reconstruct the chain and can instead aggregate local transition evidence position-agnostically.

    \item \textbf{Method.} We introduce \texttt{SeqWM}, a three-component framework consisting of history-conditioned multi-channel encoding, sliding-window detection, and random-key calibration. Together, these components embed watermark signals into conditional behavioral transitions while enabling robust verification under deletion, truncation, and partial observation.

    \item \textbf{Theory and analysis.} We analyze the robustness and statistical behavior of sequential behavioral watermarking, including the local effect of deletion under sliding-window detection and the finite-sample validity of random-key calibration without independence assumptions.

    \item \textbf{Empirical evaluation.} Across multiple agent benchmarks and open-weight LLMs, SeqWM consistently achieves reliable watermark detection while preserving agent utility, and remains substantially more robust to action-level corruption than round-indexed behavioral watermarking baselines.
\end{itemize}

\section{Related Work}
\label{sec:related_work}

\paragraph{Watermarking for Large Language Models.}
LLM watermarking is most commonly studied at the token level, where a keyed
pseudorandom rule biases generation toward a detectable subset of tokens
\citep{kirchenbauer2023watermark}. Subsequent work has improved this paradigm
through low-distortion or distribution-preserving sampling
\citep{aaronson2022watermark,christ2024undetectable,kuditipudi2024robust},
large-scale deployed systems \citep{dathathri2024scalable}, and semantic
watermarks designed to better survive paraphrasing
\citep{liu2024semantic,hou2024semstamp}. At the same time, recent studies have
clarified the limits of token-level watermarking, including reliability under
editing and length variation \citep{kirchenbauer2024reliability}, watermark
stealing \citep{jovanovic2024watermark}, robustness--utility trade-offs
\citep{pang2024nofreelunch}, and watermark inheritance through distillation
\citep{an-etal-2026-ditto}. These methods provide provenance for generated
text, but they do not directly apply to LLM agents, where the relevant evidence
is often a discrete and partially observable action trajectory rather than the
underlying token stream.

\paragraph{Watermarking for LLM Agents.}
Agentic systems interleave model decisions with environment observations and
tool executions \citep{yao2023react}, so detectors may observe only external
actions rather than internal tokens or reasoning traces. This has motivated
behavior-level watermarks for LLM agents. \textsc{Agent Guide} biases
per-step behavior choices toward keyed subsets and detects the resulting
marginal frequency shift with a statistical test
\citep{huang2025agentguide}, while \textsc{AgentMark} embeds multi-bit
identifiers into behavior choices while preserving the elicited per-step
distribution \citep{huang2026agentmark}. Other work targets trajectory or
system-level provenance: \textsc{ActHook} watermarks training trajectories via
trigger-conditioned hook actions \citep{meng2026acthook}, and
\textsc{AgentWM} protects fine-tuned agents by biasing among semantically
equivalent tool-execution paths under grey-box observation
\citep{wang2026agentwm}. These approaches establish behavioral watermarking
as a practical direction, but they largely treat evidence as per-step signals
or structured payloads. In contrast, SeqWM embeds the signal in
history-conditioned transition patterns and verifies trajectories with
position-agnostic, random-key-calibrated detection.

\section{Preliminaries}
\label{sec:prelim}

\paragraph{Sequential agent behavior.}
We consider an LLM agent that interacts with an environment over $T$ steps.
At step $t$, the agent observes a state and selects an action $b_t$ from a
finite candidate set $\mathcal{B}_t$ with $|\mathcal{B}_t| = A$.\footnote{We
take the candidate-set size $A$ to be constant across steps to simplify the
exposition. When the size varies, the same construction applies with
$A_t = |\mathcal{B}_t|$ and step-dependent null hit rate
$p_{0,t}=n/A_t$.}
Let $P_t \in \Delta(\mathcal{B}_t)$ denote the action distribution elicited from
the agent at step $t$ in the absence of any watermarking, and let
$P_t^{\mathrm{wm}}$ denote its watermarked counterpart.

\paragraph{Keyed subset primitive.}
The encoder and the detector share a secret
key $\mathcal{K}$. For any byte string $x$, both can derive a pseudorandom
subset of $\mathcal{B}_t$ via
\begin{equation}
G_{\mathcal{K}}(x) \;\triangleq\; \mathrm{SampleSubset}\!\bigl(\mathcal{B}_t,\; \mathrm{HMAC}(\mathcal{K}, x),\; n\bigr),
\label{eq:subset-primitive}
\end{equation}
where $n \le A$ is a fixed subset size. We treat
$G_{\mathcal{K}}$ as a uniformly random $n$-subset of $\mathcal{B}_t$ that is
independent across distinct contexts $x$, which is a standard idealization of
HMAC's pseudorandomness.

\paragraph{Threat model.}
The detector receives an observed action sequence $\hat{b}_{1:T'}$ together
with the secret key $\mathcal{K}$ and must decide whether the observation
originated from a watermarked source. The observation may be clean
($\hat{b}_{1:T'}=b_{1:T}$ and $T'=T$) or corrupted; in our analysis, corruption
is modeled primarily as deletion of arbitrary positions, so that $T'\le T$.

\paragraph{Notation.}
Throughout, $A_t = |\mathcal B_t|$ denotes the candidate-set size at step $t$
(and $A$ when this size is constant), $w$ the window length, $m$ the number
of channels, $\gamma$ the bias strength, and $c_t = (b_{t-w}, \ldots, b_{t-1})$
the context window at step $t$. We write $\mathcal B_{g,t}^{(j)} \subseteq
\mathcal B_t$ for the guided subset at step $t$ on channel $j$, $\|$ for
byte-string concatenation, and $\mathbbm{1}[\cdot]$ for the indicator function.

\section{Embedding Watermarks in Behavioral Transitions}
\label{sec:method}

We introduce \texttt{SeqWM}, a watermarking framework for the action
sequences produced by LLM agents. The framework is built around three
components, each addressing a distinct difficulty of watermarking sequential
agent behavior. First, a \emph{history-conditioned multi-channel encoder}
(Section~\ref{sec:encoding}) embeds the watermark signal into the agent's
conditional transition distribution rather than its marginal action frequencies,
and replicates the signal across $m$ statistically independent channels. Second,
a \emph{sliding-window detector} (Section~\ref{sec:sliding}) discards any
reliance on absolute step index, allowing detection on truncated, deleted, or
partially observed sequences without alignment information. Third, a
\emph{random-key calibration} procedure (Section~\ref{sec:calibration})
constructs an empirical null distribution from the observed sequence itself,
yielding valid statistical inference without assuming independence among
detector indicators. We begin by formalizing the setting before describing each
component in turn.

\begin{figure}[t]
    \centering
    \includegraphics[width=\linewidth]{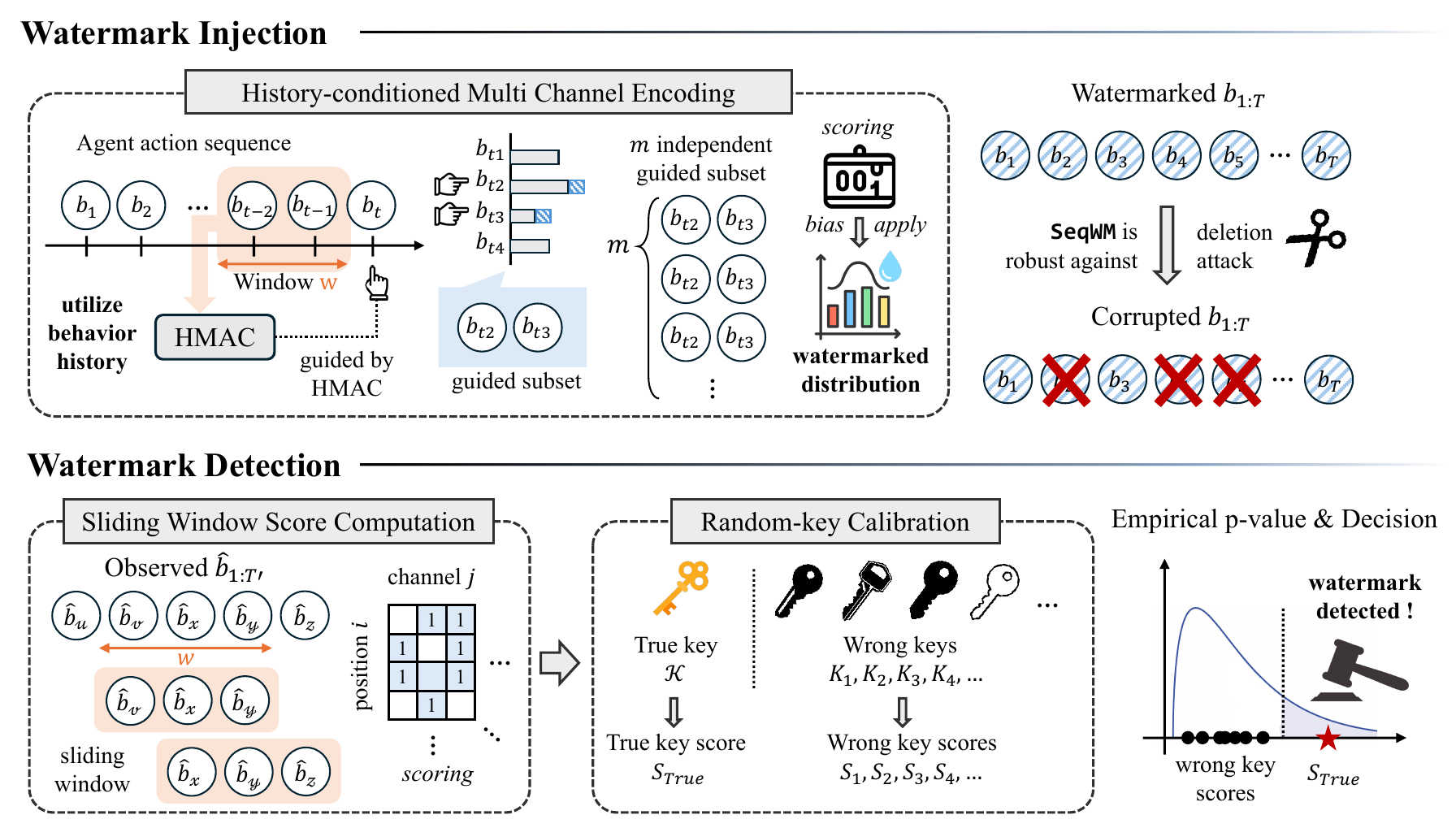}
    \caption{Overview of \texttt{SeqWM}. The injection step biases the agent's action distribution using multiple history-conditioned guided subsets, and the
detection step slides over the observed sequence and calibrates the score against
random keys to produce a p-value.}
    \label{fig:overview}
\end{figure}

\subsection{History-Conditioned Multi-Channel Encoding}
\label{sec:encoding}

\paragraph{From round indices to history-conditioned seeds.}
Round-indexed agent
watermarks (e.g., \emph{Agent Guide}) seed the guided subset directly with the
absolute step index, $B_{g,t}^{\mathrm{AG}} = G_{\mathcal{K}}(t)$, treating each
step as a statistically independent trial keyed by its position in the
trajectory. Two limitations follow. The signal is encoded into the marginal
action frequency, which is exactly what an adversary who profiles only
first-order behavior can attack; and the encoder and decoder must share an
absolute coordinate, so any deletion or truncation that perturbs the index
permanently misaligns subsequent indicators with the keyed subsets.

\texttt{SeqWM} replaces the round index with a window of preceding actions:
\begin{equation}
B_{g,t} \;=\; G_{\mathcal{K}}(c_t), \qquad c_t = (b_{t-w}, \ldots, b_{t-1}).
\label{eq:history-seed}
\end{equation}
This seemingly small change has two structural consequences.
First, the watermark signal is now carried by the conditional transition
distribution $P_t(\cdot \mid c_t)$: marginal action frequencies are not biased
upward, only the dependence of $b_t$ on its recent context is. An adversary who
matches the marginal distribution therefore cannot remove the watermark without
also disrupting transition statistics. Second, since the same window
deterministically yields the same subset wherever it appears in the stream, the
encoder and the decoder no longer require a shared notion of absolute
position---a property we exploit in Section~\ref{sec:sliding}.

\paragraph{Multi-channel redundancy.}
We further replicate the watermark across $m$ statistically independent channels by appending a channel index to the seed:
\begin{equation}
B_{g,t}^{(j)} \;=\; G_{\mathcal{K}}\!\bigl(c_t \,\|\, j\bigr), \qquad j = 1, \ldots, m.
\label{eq:multi-channel}
\end{equation}
The pseudorandomness of $\mathrm{HMAC}$ ensures that the $m$ subsets behave as independent random $n$-subsets.
We then summarize how strongly each candidate action is favored by the watermark via the multi-channel score
\begin{equation}
s_t(b) \;=\; \sum_{j=1}^{m} \mathbbm{1}\!\bigl[b \in B_{g,t}^{(j)}\bigr] \;\in\; \{0, 1, \ldots, m\},
\label{eq:score}
\end{equation}
which counts the number of channels in which $b$ is a watermark
target. The watermarked distribution applies an exponential tilt to the elicited
distribution in proportion to this score:
\begin{equation}
P_t^{\mathrm{wm}}(b) \;=\; \frac{1}{Z_t}\, P_t(b)\, \exp\!\left(\gamma \cdot \frac{s_t(b)}{m}\right),
\qquad Z_t \;=\; \sum_{b' \in \mathcal{B}_t} P_t(b')\, \exp\!\left(\gamma \cdot \frac{s_t(b')}{m}\right).
\label{eq:tilt}
\end{equation}

The exponential form is chosen for three reasons. It places the
encoder in an exponential family with sufficient statistic $s_t(b)/m$ and
natural parameter $\gamma$, which makes the subsequent KL analysis transparent.
For small $\gamma$ it reduces to a multiplicative bias, since $\exp(\gamma s/m)
\approx 1 + \gamma s/m$. And it naturally amplifies actions that are favored by
multiple channels relative to those favored by only one.

The encoder draws $b_t \sim P_t^{\mathrm{wm}}$ for $t > w$ and falls back to the
unmodified $P_t$ during the first $w$ bootstrap steps when no full window is yet
available. The procedure is summarized in Algorithm~\ref{alg:encoder}.

\paragraph{Detection power and utility cost.}
We characterize the encoder's
per-step effect under a sum-score detector. Let $p_0 = n/A$ denote the
probability that an action drawn from $P_t$ falls into a single guided subset,
which holds for any one channel under the pseudorandomness of $G_{\mathcal{K}}$.
A first-order expansion of the exponential tilt around $\gamma = 0$, combined
with channel independence, gives the corresponding probability under the
watermarked distribution as
\begin{equation}
p_1 \;=\; p_0 + \frac{\gamma}{m}\, p_0(1 - p_0) + O(\gamma^2).
\label{eq:p1}
\end{equation}
A sum-score detector that aggregates $N = (T - w)\, m$ indicators across steps and channels
therefore produces an expected mean shift $\mu_1 - \mu_0 = N(p_1 - p_0) = (T - w)\, \gamma\, p_0(1 - p_0)$.
Dividing by the binomial standard deviation $\sqrt{N\, p_0(1 - p_0)}$ yields the signal-to-noise ratio
\begin{equation}
\mathrm{SNR} \;=\; \sqrt{\frac{T - w}{m}}\, \cdot\, \gamma \sqrt{p_0(1 - p_0)}.
\label{eq:snr}
\end{equation}

On the utility side, the second-order expansion of KL divergence
in an exponential family, together with $s_t(b) \sim \mathrm{Binomial}(m, p_0)$
under uniform $P_t$, gives
\begin{equation}
D_{\mathrm{KL}}\!\bigl(P_t^{\mathrm{wm}} \,\big\|\, P_t\bigr)
\;=\; \frac{\gamma^2}{2}\, \mathrm{Var}_{b \sim P_t}\!\left[\frac{s_t(b)}{m}\right] + O(\gamma^3)
\;=\; \frac{\gamma^2}{2 m}\, p_0(1 - p_0) + O(\gamma^3).
\label{eq:kl}
\end{equation}

\begin{wraptable}{r}{0.45\textwidth}
\centering
\caption{Mean-shift validation. The measured score gap $\Delta S$ matches
the first-order theoretical prediction $\Delta S_{\text{pred}}$ within a 
small constant factor across all three domains.}
\label{tab:effect-size}
\vspace{0.7em}
\begin{tabular*}{\linewidth}{@{\extracolsep{\fill}}lrrr@{}}
\toprule
Dataset & $\Delta S_{\text{pred}}$ & $\Delta S$ & Ratio \\
\midrule
ToolBench & 193.0 & 216.8 & 1.12 \\
ALFWorld  & 348.5 & 318.1 & 0.91 \\
OASIS     &  43.8 &  62.4 & 1.42 \\
\bottomrule
\end{tabular*}
\vspace{-1em}
\end{wraptable}

The mean-shift prediction in Eq.~\eqref{eq:snr}'s numerator is directly
testable: a watermarked sequence should produce $S - \mathbb{E}[S] \approx
(T-w)\gamma p_0(1-p_0)$. Table~\ref{tab:effect-size} confirms this across
all three domains: the measured gap matches the first-order prediction
within a small constant factor (ratios $0.91$ to $1.42$), with deviations
consistent with the $O(\gamma^2)$ correction at $\gamma=2.0$. Random-key
calibration also yields an unbiased null estimator: under unwatermarked
sequences, the measured gap is at most $1.74\%$ of $\mathbb{E}[S]$ across
all domains.

Eqs.~\eqref{eq:snr} and \eqref{eq:kl} jointly imply that the ratio
$\mathrm{SNR}^2 / D_{\mathrm{KL}}$ is independent of $m$. Under a plain
sum-score detector, redundant channels do not improve the raw
information-theoretic trade-off between detection power and distortion---a fact
we state plainly because it is easy to obscure. The value of $m$ instead lies on
three orthogonal axes that the SNR/KL ratio does not capture.
\emph{(i) Robustness buffer.} Redundant channels provide multiple local indicators for
each transition, so deletions remove evidence locally rather than causing the
global misalignment suffered by round-indexed detectors. Under the normalized
tilt in Eq.~(\eqref{eq:tilt}), however, the sum-score mean shift does not grow linearly with
$m$; the robustness benefit is therefore a local redundancy effect rather than
an information-theoretic gain for the plain sum-score detector.
\emph{(ii) Calibration concentration.} The
relative standard deviation of the wrong-key score used in
Section~\ref{sec:calibration} scales as $\bigl((T - w)\, m\bigr)^{-1/2}$,
sharpening the empirical null distribution and improving p-value resolution at
modest sequence length and calibration size.
\emph{(iii) Detector extensibility.} The multi-valued score $s_t(b) \in \{0, \ldots, m\}$ admits
weighted detectors and likelihood-ratio tests that strictly dominate the
sum-score; the latter is the simplest, not the most powerful, use of the
multi-channel structure, and the calibration procedure of
Section~\ref{sec:calibration} accommodates any such replacement without
modification.

\subsection{Sliding-Window Detection}
\label{sec:sliding}

\paragraph{Removing absolute position from detection.}
A round-indexed detector verifies, for each $t$, whether $b_t \in
G_{\mathcal{K}}(t)$, and is therefore brittle: a single deletion shifts every
downstream action by one index, so $b_{t+1}$ is now compared against a subset
that was never used to bias it, and the resulting noise propagates indefinitely.
The history-conditioned encoder of Section~\ref{sec:encoding} permits a
structurally different detector that discards the absolute index entirely. Since
the same window deterministically yields the same guided subsets under the
encoder and the decoder, evidence of watermarking can be aggregated directly
from every length-$w$ substring of the observed sequence:
\begin{equation}
S(\mathcal{K}, \hat{b}_{1:T'})
\;=\; \sum_{i=1}^{T' - w} \sum_{j=1}^{m}
\mathbbm{1}\!\left[\hat{b}_{i + w} \in G_{\mathcal{K}}\!\bigl(\hat{b}_{i:i + w - 1} \,\|\, j\bigr)\right].
\label{eq:detect-score}
\end{equation}

Each term examines whether the action immediately following a window matches the
guided subset that the encoder would have used at \emph{some} step where the
same window appeared. The detector neither knows nor needs to know which step
that was. Algorithm~\ref{alg:detector-score} summarizes the score computation.

\paragraph{An asymmetry between encoder and decoder.}
The construction exploits an asymmetry that does not arise in conventional
per-step watermarks. The encoder is naturally chained: each step's bias depends
on the immediately preceding $w$ actions. The decoder, however, treats every
length-$(w+1)$ substring of the observation as an independent piece of evidence
and aggregates over them without following the chain. The situation is analogous
to message authentication codes, where a single key permits sequential
generation but allows verification of arbitrary fragments of a stream. This
asymmetry is what enables robust detection on partial or rearranged
observations.

\paragraph{Robustness to deletion.}
The structural difference between round-indexed and sliding-window detection
manifests as a sharp asymptotic separation under deletion.




\begin{theorem}[Deletion robustness]
\label{thm:deletion}
Let $b_{1:T}$ be a watermarked sequence and let $\hat{b}$ be obtained by
deleting $d$ positions chosen uniformly at random, with deletion fraction
$\rho = d / T$. Assume the verifier can reconstruct the candidate action set
used for each scored transition.

We analyze the round-indexed baseline under the standard pseudorandom
misalignment idealization: once a deletion shifts the detector's round index,
an action generated under $G_{\mathcal K}(t')$ and tested against
$G_{\mathcal K}(t)$ for $t\neq t'$ contributes only the null hit rate $p_0$.

\emph{(a) Round-indexed detection.}
Let \(k^*\) denote the
position of the first deletion. For all detector positions after the aligned
prefix, i.e., \(t \ge k^*\), the round-indexed indicator
\(\mathbbm{1}[\hat b_t \in G_{\mathcal K}(t)]\) has expectation \(p_0\).
Consequently, the expected score gap of a round-indexed detector satisfies
\begin{equation}
\mathbb{E}\bigl[S^{\mathrm{RI}}_{\mathrm{after}} - S^{\mathrm{RI}}_{\mathrm{null}}\bigr]
\;=\; O(1/\rho),
\label{eq:ri-collapse}
\end{equation}
which vanishes relative to the clean-sequence gap $\Theta(T)$ as $T \to \infty$
for any fixed $\rho > 0$.

\emph{(b) Sliding-window detection.}
The score $S(\mathcal{K}, \hat{b})$ defined in Eq.~\eqref{eq:detect-score}
satisfies the deterministic additive bound
\begin{equation}
S(\mathcal{K}, \hat{b})
\ge
S(\mathcal{K}, b_{1:T}) - d(w+1)m .
\end{equation}
Equivalently, the number of clean sliding-window indicators that can be
invalidated by the deletions is at most $d(w+1)m$, independent of $T$.
\end{theorem}

The proof of (a) is a direct consequence of round-index misalignment: once
position $k^*$ is deleted, the encoder used $G_{\mathcal{K}}(t)$ to bias the
action at original step $t$, but the detector now reads it as step $t - 1$ and
compares against $G_{\mathcal{K}}(t - 1)$, which is an independent random
subset under the pseudorandomness of $G$. The proof of (b) is the counting
argument outlined above: each deleted position appears in at most $w$ context
windows and is the next-action of at most one window, so it invalidates at
most $(w + 1)\, m$ indicators in Eq.~\eqref{eq:detect-score}. Full proofs
are deferred to Appendix~\ref{app:proofs}.

Theorem~\ref{thm:deletion} captures the central design contrast of our method.
For example, when $w=3$, deleting a fraction $\rho$ of the trajectory can
invalidate at most $4\rho T m$ clean sliding-window indicators. This contrasts
with round-indexed detection, where the expected aligned prefix length is only
$O(1/\rho)$. The bound is additive rather than a direct guarantee on the
post-deletion detection gap; the empirical results in Section~5 measure the
resulting detection power.

\subsection{Random-Key Calibration}
\label{sec:calibration}

\paragraph{Why a closed-form null distribution does not apply.}
The natural inferential procedure for the score in Eq.~\eqref{eq:detect-score}
would be a z-test against a binomial null with variance $N\, p_0(1 - p_0)$. This
null is, however, invalid in our setting. Three sources of dependence couple the
indicators that the test would treat as independent: actions $b_t$ depend on
$b_{t-1}$ through the LLM's natural sequential structure; the watermark itself
reuses the same guided subset whenever the same window reappears; and the
overlapping length-$(w+1)$ substrings examined by the sliding-window detector
share actions by construction. Each of these typically induces positive
correlation, so a binomial-based test underestimates the variance of $S$ under
the null and inflates false-positive rates---an unacceptable failure mode for a
statistical detector.

\paragraph{The wrong-key trick.}
We sidestep the closed-form null by constructing the null distribution
\emph{empirically from the observed sequence itself}. The construction rests on
a single observation: re-evaluating the score in Eq.~\eqref{eq:detect-score} on
the same sequence $\hat{b}_{1:T'}$ but with a wrong key $\mathcal{K}_r \ne
\mathcal{K}$ preserves \emph{every} dependence in the sequence---both natural
and watermark-induced---while erasing only the keyed alignment that produces the
watermark signal. Whatever positive correlation the binomial null fails to
account for is therefore present in equal measure in the wrong-key scores, and
is automatically absorbed into the empirical null.

Concretely, we draw $M$ independent random keys $\mathcal{K}_1, \ldots,
\mathcal{K}_M$ from the key distribution, compute $S_r = S(\mathcal{K}_r,
\hat{b}_{1:T'})$ under each, and report the empirical p-value
\begin{equation}
p \;=\; \frac{1 + \bigl|\{r : S_r \ge S_{\mathrm{true}}\}\bigr|}{M + 1},
\qquad S_{\mathrm{true}} = S(\mathcal{K}, \hat{b}_{1:T'}),
\label{eq:pvalue}
\end{equation}
declaring the sequence watermarked iff $p < \alpha$. The full procedure is given in Algorithm~\ref{alg:calibration}.

\paragraph{Validity and finite-sample guarantee.}
Under the null hypothesis that $\hat{b}_{1:T'}$ was not generated using
$\mathcal{K}$, the sequence is statistically independent of every key, and the
true and wrong keys are exchangeable draws from the key distribution. The vector
$(S_{\mathrm{true}}, S_1, \ldots, S_M)$ is therefore exchangeable, the rank of
$S_{\mathrm{true}}$ is uniform on $\{1, \ldots, M + 1\}$, and the p-value in
Eq.~\eqref{eq:pvalue} satisfies $\Pr[p \le \alpha] \le \alpha$ for every $\alpha
\in (0, 1)$. The guarantee is finite-sample: it does not invoke any asymptotic
regime in $T'$, $m$, or $M$. Under the alternative, the true-key score
$S_{\mathrm{true}}$ is systematically larger than the wrong-key scores, and the
gap grows in $\gamma$ and in sequence length.

\paragraph{Three properties of the resulting test.}
The wrong-key trick yields a test with several practical consequences.
\emph{(i) Distribution-free.} Any heterogeneity, skew, or temporal structure in
$P_t$ is mirrored in every $S_r$ and cancels in the comparison; no assumption on
the agent's action distribution enters the validity argument.
\emph{(ii) Robust to adversarial corruption.} If the sequence has been deleted,
truncated, substituted, or partially observed, the same corrupted sequence is
presented to every key; the corruption affects $S_{\mathrm{true}}$ and $\{S_r\}$
symmetrically and the p-value remains valid, though detection power may decrease
as the watermark signal weakens.
\emph{(iii) Detector-agnostic.} Replacing Eq.~\eqref{eq:detect-score} with any
other test statistic computed from $(\mathcal{K}, \hat{b}_{1:T'})$---a
likelihood ratio, a weighted aggregator, or a learned scorer---preserves
validity, providing a principled route to stronger detectors that exploit the
multi-channel structure of Section~\ref{sec:encoding}.

\section{Experiments and Analysis}
\label{sec:experiments}

\subsection{Experimental Setup}
\label{subsec:setup}
We evaluate our method on three agent benchmarks: ToolBench (200 queries from
under a ReAct loop with at most 6 steps per episode), ALFWorld (50 tasks with
maximum episode length of 50 and one-shot prompting), and OASIS Reddit (100
simulation steps with one user agent and three NPC agents). The agent LLM is one
of LLaMA-3.2 3B Instruct, Gemma 4 E4B-it, or Qwen3-4B Instruct, decoded at
temperature 0.7. Unless otherwise noted, watermark parameters are fixed at
window size $w=3$, $m=8$ channels, guided-subset size $n=3$ (with $n=2$ on OASIS
due to its smaller behavior set), floor $\delta=0.2$, bias strength
$\gamma=2.0$, and a detection threshold of $\alpha=0.01$ with $M=1000$
random-key calibration trials. The $\gamma$-sensitivity ablation sweeps $\gamma
\in \{0.5, 1.0, 1.5, 2.0, 2.5, 3.0\}$ with $m$ fixed at 8 on a 50-query
ToolBench subset. All main rollouts run on NVIDIA A100 GPUs, with sanity checks
and analysis sweeps offloaded to NVIDIA RTX PRO 6000 Blackwell GPUs on the same
SLURM cluster. Full configuration details are deferred to
Appendix~\ref{app:setup}.

\subsection{Detection Results}
\label{subsec:detection_results}

\begin{table}[t]
\centering
\caption{Watermark detection across three benchmarks and three LLMs. We report the detection $z$-score, $p$-value, and per-step hit rate (Hit\%, undefined for unwatermarked and AgentMark). \texttt{SeqWM} is the only method that reliably rejects the null across all settings.}
\label{tab:wm_results}
\setlength{\tabcolsep}{4pt}
\renewcommand{\arraystretch}{1.1}
\newcommand{\hl}{\cellcolor{gray!15}}
\begin{tabular}{c|l|ccc|ccc|ccc}
\toprule
\multirow{2}{*}{\textbf{Model}} & \multicolumn{1}{c|}{\multirow{2}{*}{\textbf{Method}}}
& \multicolumn{3}{c|}{\textbf{ToolBench}}
& \multicolumn{3}{c|}{\textbf{ALFWorld}}
& \multicolumn{3}{c}{\textbf{OASIS}} \\
& & \textbf{$z \uparrow$} & \textbf{$p \downarrow$} & \textbf{Hit\%}
& \textbf{$z \uparrow$} & \textbf{$p \downarrow$} & \textbf{Hit\%}
& \textbf{$z \uparrow$} & \textbf{$p \downarrow$} & \textbf{Hit\%} \\
\midrule
\multirow{4}{*}{Gemma}
& Unwatermarked   &  0.13 & 0.455 & --   & -0.93 & 0.830 & --   & -0.14 & 0.571 & --   \\
& AgentGuide      &  1.83 & 0.032 & 23\% &  0.89 & 0.197 & 17\% & -1.06 & 0.861 & 49\% \\
& AgentMark$^\dagger$       & -0.71 & 0.761 & --   & -1.80 & 0.965 & --   &  1.09 & 0.152 & --   \\
& \hl\texttt{SeqWM} \textit{(ours)} & \hl\textbf{7.77} & \hl\textbf{0.001} & \hl\textbf{67\%}
                           & \hl\textbf{9.45} & \hl\textbf{0.001} & \hl\textbf{66\%}
                           & \hl\textbf{4.67} & \hl\textbf{0.001} & \hl\textbf{99\%}  \\
\midrule
\multirow{4}{*}{LLaMA}
& Unwatermarked   & -0.21 & 0.580 & --   & -0.36 & 0.645 & --   & -0.24 & 0.609 & --   \\
& AgentGuide      & -0.54 & 0.707 & 33\% & -0.21 & 0.598 & 20\% & -1.06 & 0.861 & 49\% \\
& AgentMark$^\dagger$       & -1.97 & 0.981 & --   &  0.67 & 0.253 & --   &  1.09 & 0.152 & --   \\
& \hl\texttt{SeqWM} \textit{(ours)} & \hl\textbf{9.22}  & \hl\textbf{0.001} & \hl\textbf{81\%}
                           & \hl\textbf{10.18} & \hl\textbf{0.001} & \hl\textbf{75\%}
                           & \hl\textbf{5.16}  & \hl\textbf{0.001} & \hl\textbf{100\%} \\
\midrule
\multirow{4}{*}{Qwen}
& Unwatermarked   &  0.08 & 0.466 & --   &  0.21 & 0.416 & --   &  0.15 & 0.455 & --   \\
& AgentGuide      & -1.47 & 0.928 & 23\% & -2.17 & 0.985 & 20\% &  1.01 & 0.165 & 44\% \\
& AgentMark$^\dagger$       & -1.14 & 0.879 & --   &  1.34 & 0.087 & --   & -1.03 & 0.857 & --   \\
& \hl\texttt{SeqWM} \textit{(ours)} & \hl\textbf{5.48}  & \hl\textbf{0.001} & \hl\textbf{63\%}
                           & \hl\textbf{11.44} & \hl\textbf{0.001} & \hl\textbf{79\%}
                           & \hl\textbf{2.53}  & \hl\textbf{0.010} & \hl\textbf{97\%} \\
\bottomrule
\end{tabular}
\end{table}
\footnotetext{$\dagger$~AgentMark is a multi-bit payload watermark, so the single-bit detection metrics ($z$, $p$) are not its native evaluation; we include them for cross-method comparability. Hit\% is undefined under its differential bin encoding.}

Table~\ref{tab:wm_results} shows that \texttt{SeqWM} consistently achieves
strong watermark detection across all three benchmarks and all evaluated LLMs,
while the baselines frequently fail to separate from the null.

A notable pattern is that strong sequence-level detection emerges even when the
per-step hit rates are only moderately elevated. This indicates that the
watermark is not concentrated in a small number of conspicuous actions, but
instead accumulates gradually through many weak transition-level biases across
the trajectory. The results therefore support the central premise of our method:
sequential dependence itself can serve as a robust substrate for behavioral
watermarking.

\begin{figure}
    \centering
    \includegraphics[width=\linewidth]{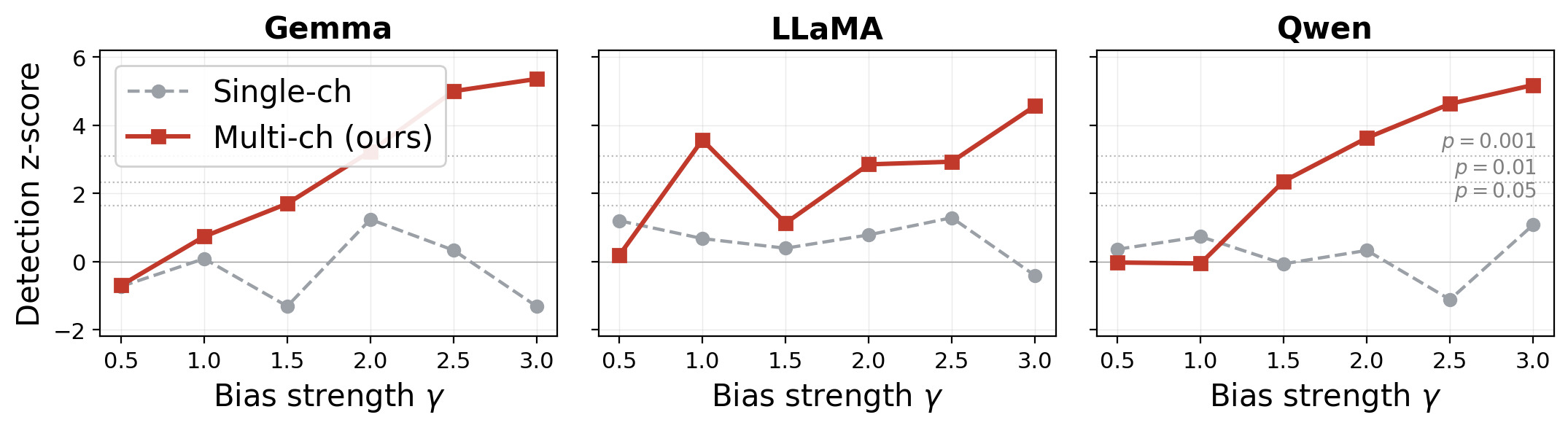}
    \caption{Detection z-score versus bias strength on OASIS Reddit across
three LLM backbones; dotted lines mark standard significance thresholds. Multi-channel detection scales monotonically with bias and clears the
strictest threshold on every backbone, while the single-channel
ablation stays at the noise floor throughout.}

    \label{fig:gamma}
\end{figure}

\paragraph{$\gamma$ Sensitivity.}
We sweep $\gamma \in \{0.5,1.0,1.5,2.0,2.5,3.0\}$ on OASIS across three LLM
backbones. We use OASIS as a challenging stress test because its trajectories are
relatively short and its action vocabulary is small, leaving less room for
watermark evidence to accumulate. Figure~\ref{fig:gamma} shows that the
multi-channel detector strengthens steadily as $\gamma$ increases and reaches
$p<0.01$ from $\gamma=2.0$ onward on all three backbones, suggesting a stable
operating range rather than a model-specific tuning artifact. The single-channel
ablation, however, remains near the null even at $\gamma=3.0$. Thus, detection
does not come from bias strength alone: reliable detection requires redundant
evidence accumulated across channels together with the position-agnostic
sequential detector. We therefore set $\gamma=2.0$ as the default for the
remaining experiments, choosing the smallest value that consistently yields
reliable detection across backbones.

\subsection{Robustness}
\label{subsec:attack}

We evaluate robustness under random action deletion, where a fraction
$r \in \{0,5,10,20,30,50\}\%$ of actions is removed before detection. This setting
captures both partial logging and adversarial attempts to break trajectory-level
provenance.

\begin{figure}
    \centering
    \includegraphics[width=\linewidth]{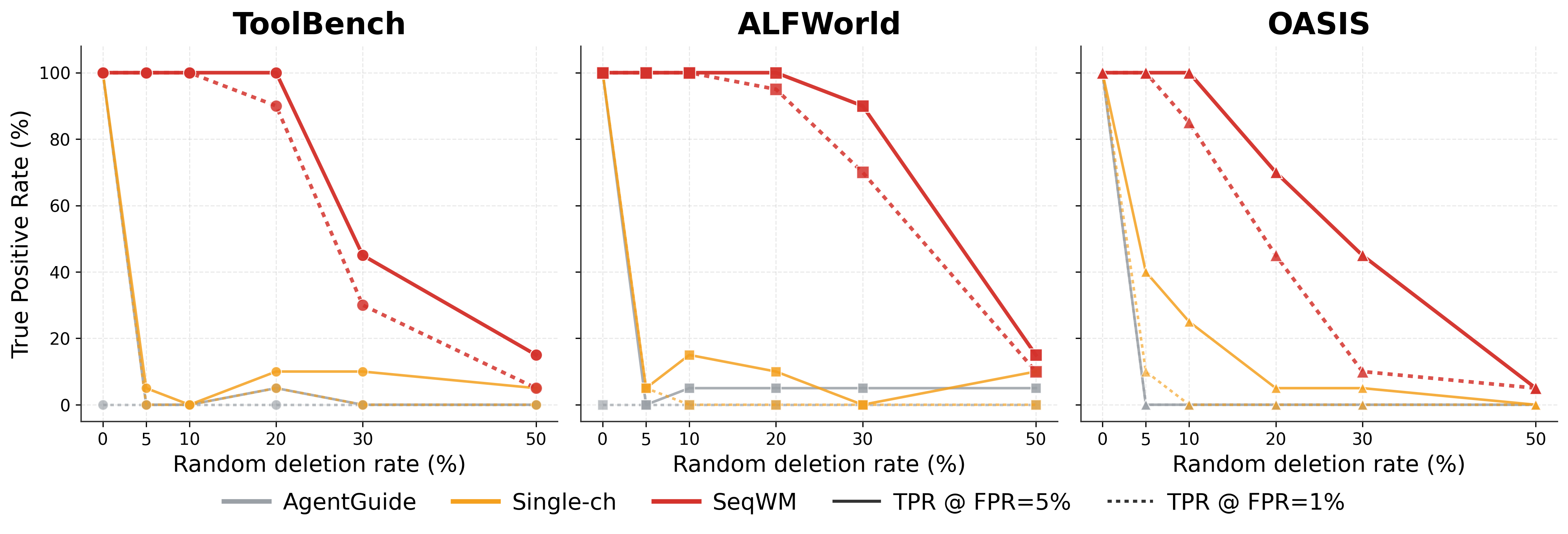}
    \caption{Detection TPR as a function of random action deletion rate across three domains (ToolBench, ALFWorld, OASIS). Colors distinguish methods and line styles distinguish FPR thresholds. Higher curves indicate stronger robustness to deletion attacks.}
    \label{fig:attack}
\end{figure}

Figure~\ref{fig:attack} shows that round-indexed detection is brittle:
AgentGuide collapses after small deletion rates because a single missing action
misaligns all downstream keyed subsets. The single-channel variant avoids this
global misalignment, but its score margin is too thin to survive substantial
corruption.

In contrast, the full multi-channel SeqWM detector remains stable under moderate
deletion across all three domains. Detection stays near-perfect through small
deletion rates and degrades gradually rather than catastrophically as the
deletion rate increases. This supports the design of SeqWM: history-conditioned
seeding localizes the effect of deletion, while multi-channel redundancy
preserves enough evidence from the surviving contextual transitions.

\section{Conclusion}
\label{sec:conclusion}
We presented \texttt{SeqWM}, a sequential behavioral watermarking framework for LLM agents that treats provenance as evidence accumulated from how actions evolve over time, rather than from marginal per-step action frequencies. By conditioning watermark signals on recent action history, distributing them across multiple behavioral channels, and verifying trajectories with position-agnostic sliding-window detection calibrated by random keys, \texttt{SeqWM} removes the need for fragile round alignment while preserving statistically valid detection. This design makes trajectory corruption a local source of evidence loss rather than a global detector failure: deletions, truncation, and partial observation affect only the surrounding transition windows instead of misaligning the remainder of the sequence. Across diverse agent benchmarks and LLM backbones, \texttt{SeqWM} achieves reliable watermark detection with preserved utility and remains robust under deletion rates that cause round-indexed and bit-encoding baselines to collapse. More broadly, our results suggest that sequential structure is not merely a nuisance in agent watermarking, but a powerful carrier of provenance signals; extending this perspective to stronger adaptive attacks, substitutions, and semantic paraphrase-style trajectory edits is an important direction for future work.









\bibliographystyle{plainnat}  
\bibliography{references}   

\appendix

\section{Social Impact and Limitations}
\label{sec:limitations}

Reliable provenance for autonomous LLM agents is an increasingly important capability as such systems are deployed in consequential tool-use settings, and we view action-trace watermarking as net-positive for accountability and post-hoc auditing; deployers should nonetheless communicate clearly when such schemes are active. Several natural extensions remain open and we view them as promising directions rather than shortcomings of the present construction. Stronger adversarial models, including adaptive attackers and semantic-level transformations such as paraphrasing, raise questions that any behavior-level watermark must eventually engage with, and we expect the redundant-coding perspective developed here to provide a useful starting point. Tightening the analytical characterization of multi-channel redundancy beyond the orthogonal benefits we identify (calibration concentration, robustness buffer, detector extensibility) is likewise an interesting theoretical direction.

\section{Full Related Work}
\label{app:full_related_work}

\subsection{Watermarking for Large Language Models}

Watermarking has emerged as a principled mechanism for attributing content
generated by large language models (LLMs). The seminal scheme of
\citet{kirchenbauer2023watermark} partitions the vocabulary into green and
red lists via a keyed pseudorandom function and biases the sampling
distribution toward the green list, enabling zero-bit detection with a
$z$-statistic. Subsequent work has sought to reduce the utility or
perceptual cost of biased sampling. Low-distortion and
distribution-preserving designs include the Gumbel-style scheme of
\citet{aaronson2022watermark}, the cryptographically undetectable
construction of \citet{christ2024undetectable}, the inverse-transform
sampling watermark of \citet{kuditipudi2024robust}, and the production-ready
SynthID-Text system of \citet{dathathri2024scalable}, which was evaluated in
a live deployment involving nearly twenty million Gemini responses.
A complementary line conditions the watermark on sentence-level semantics to
improve robustness to paraphrasing and lexical rewriting
\citep{liu2024semantic,hou2024semstamp}. Cross-lingual robustness has been
studied separately: translation can substantially weaken existing text
watermarks, motivating cross-lingual defenses such as X-SIR
\citep{he-etal-2024-watermarks}. Orthogonal refinements balance watermark
mass across token-rank partitions \citep{park2026watermod} or adapt the
green-list bias to local syntactic predictability
\citep{park2026linguistics}.

A growing body of work probes the limits of this paradigm.
\citet{kirchenbauer2024reliability} stress-test detection under paraphrase
and length variation; \citet{jovanovic2024watermark} demonstrate that
current schemes admit watermark stealing, in which an adversary approximates
secret watermarking rules from observed outputs and then mounts both
spoofing and scrubbing attacks; and \citet{pang2024nofreelunch} analyze
no-free-lunch trade-offs among robustness, utility or usability, and
resistance to removal and spoofing attacks. More recently,
\citet{an-etal-2026-ditto} show that a student model trained on outputs from
a watermarked victim can inherit and reuse the watermark signal, enabling
spoofing without access to the victim's logits, secret key, or watermarking
scheme.

A separate strand operates on the model or training data rather than only on
sampled outputs. Model fingerprinting embeds keyed challenge--response
associations: instructional fingerprinting uses lightweight instruction
tuning so that private triggers elicit identifying responses
\citep{xu2024instructional}, while Chain \& Hash constructs hash-linked
question--answer challenges for model ownership verification
\citep{russinovich2026chainhash}. Semantically conditioned variants further
attach the fingerprint to broad semantic domains, reducing reliance on small
fixed trigger sets and making simple trigger filtering or purification less
effective \citep{gloaguen2026semantic}. The closest analogues for
watermarking training corpora are code dataset watermarks, which introduce
data-poisoning or backdoor-style features \citep{sun2022coprotector} or
adaptive semantics-preserving program transformations
\citep{sun2023codemark}; such schemes have nevertheless been shown to be
detectable and removable through automated code abstraction
\citep{xiao2025decoma}. Across these axes, however, the evidence used for
verification is generated text, challenge--response behavior, code artifacts,
or model-level behavior. These settings do not directly address provenance
for the discrete, partially observable action trajectories produced by LLM
agents.

\subsection{Watermarking for LLM Agents}

The schemes above generally assume that detection can inspect the relevant
token sequence, challenge-response interface, or model artifact. Modern LLM
agents violate this assumption. Agent execution interleaves model-generated
reasoning or actions with environment-generated observations
\citep{yao2023react}, and in many deployed or grey-box agentic settings,
internal token streams and reasoning traces may be unavailable, leaving only
visible actions and final responses as practical evidence
\citep{huang2025agentguide,huang2026agentmark,wang2026agentwm}. Token-level
signals can be diluted by non-trainable observations and are largely lost
when high-level decisions are compiled into low-bandwidth tool invocations
such as \texttt{bookmark} or \texttt{send\_email}. These properties have
motivated a separate line of work that watermarks agent behavior rather than
surface text.

\paragraph{Behavior-level watermarking.}
\citet{huang2025agentguide} introduce \textsc{Agent Guide}, a behavioral
watermarking framework for LLM agents. The scheme elicits a probability
distribution over a finite list of high-level behaviors at each step and
applies watermark-guided probability biases to a keyed subset, so that
watermarked agents select guided behaviors more frequently than expected
under the unwatermarked null. Detection is performed over multiple decision
rounds using a one-sided $z$-test against a binomial null. Because the scheme
explicitly changes per-step behavior probabilities, however, its
distributional cost can accumulate over long horizons. Building on this
limitation, \citet{huang2026agentmark} propose \textsc{AgentMark}, which
embeds multi-bit identifiers into the same behavioral layer while preserving
the elicited per-step distribution. Their concrete instantiation combines
distribution recombination of behavior probabilities into uniform bins with
cyclic-shift uniform encoding within each bin
\citep{kaptchuk2021meteor,ding2023discop,liao2025fdpss}, together with
random linear network coding for erasure-resilient recovery from partial
trajectories. The two schemes therefore place the statistical signal in
different locations: \textsc{Agent Guide} concentrates evidence in marginal
frequency shifts of guided behaviors and admits a simple $z$-test, whereas
\textsc{AgentMark} preserves the per-step marginal distribution and recovers
a sequence-level payload through structured decoding.

\paragraph{Trajectory and system-level provenance.}
A complementary thread targets the training data and the deployed system
rather than runtime decision sampling. \citet{meng2026acthook} address the
unauthorized redistribution of expert or high-value agent trajectories
\citep{pan2025swegym,yang2025swesmith,xu2025agenttrek,li2023apibank,gou2025mind2web2},
whose construction and validation can require substantial annotation and
labeling effort \citep{bhatia2025spice}. Their method, \textsc{ActHook},
injects auxiliary hook actions at ambiguous or high-entropy decision points,
conditioned on a secret activation phrase appended to a small subset of
training prompts. An agent fine-tuned on the watermarked corpus learns this
trigger-conditioned behavior, which is then verified through a paired
hypothesis test against a sham trigger. 
\citet{wang2026agentwm} take a third
perspective and treat the entire fine-tuned agent as the proprietary asset
under a grey-box visibility model in which only the action sequence and final
response are observable. In this setting, \textsc{AgentWM} exploits
semantically equivalent action sequences by biasing the distribution over
functionally identical tool execution paths, with verification performed via
statistical hypothesis testing.

Beyond direct watermarking, a growing body of work documents the broader IP
attack surface of agentic systems, including prompt extraction in
multi-agent pipelines \citep{wang2025ipleakage}, malicious code execution in
multi-agent orchestration \citep{triedman2025multiagent}, systematic surveys
of agent security risks \citep{deng2025aiagents}, and the broader
model-extraction threat landscape for LLMs \citep{zhao2025measurvey}. These
threats are largely orthogonal to the watermarking question, yet they
reinforce the case for auditable provenance mechanisms in regulated
deployments.

\section{Algorithm Pseudocode}
\label{app:algorithms}
We provide explicit pseudocode for the three procedures of \texttt{SeqWM}:
the multi-channel encoder (Algorithm~\ref{alg:encoder}), the sliding-window
score computation (Algorithm~\ref{alg:detector-score}), and the random-key
calibrated detector (Algorithm~\ref{alg:calibration}).
\begin{algorithm}[h!]
\caption{Multi-channel encoder.}
\label{alg:encoder}
\begin{algorithmic}[1]
\Require key $\mathcal{K}$, horizon $T$, window $w$, channels $m$, subset size $n$, bias $\gamma$
\State $\mathrm{history} \gets [\,]$
\For{$t = 1, \ldots, T$}
    \State $(\mathcal{B}_t, P_t) \gets \textsc{ElicitFromLLM}(\text{state}_t)$
    \If{$|\mathrm{history}| < w$}
        \State $b_t \sim P_t$ \Comment{bootstrap step: no watermark}
    \Else
        \State $c_t \gets \mathrm{history}[-w{:}\,]$
        \State $B_{g,t}^{(j)} \gets G_{\mathcal{K}}(c_t \,\|\, j)$ for $j = 1, \ldots, m$
        \State Compute $s_t(b)$ and $P_t^{\mathrm{wm}}(b)$ for all $b \in \mathcal{B}_t$ via Eqs.~\eqref{eq:score}--\eqref{eq:tilt}
        \State $b_t \sim P_t^{\mathrm{wm}}$
    \EndIf
    \State $\mathrm{history}.\textsc{append}(b_t)$
\EndFor
\State \Return $b_{1:T}$
\end{algorithmic}
\end{algorithm}

\begin{algorithm}[h!]
\caption{Sliding-window score computation.}
\label{alg:detector-score}
\begin{algorithmic}[1]
\Require key $\mathcal{K}$, observed sequence $\hat{b}_{1:T'}$, parameters $(w, m, n)$
\State $S \gets 0$
\For{$i = 1, \ldots, T' - w$}
    \State $c \gets \hat{b}_{i : i + w - 1}$, \quad $b_{\mathrm{next}} \gets \hat{b}_{i + w}$
    \For{$j = 1, \ldots, m$}
        \State $B_g \gets G_{\mathcal{K}}(c \,\|\, j)$
        \If{$b_{\mathrm{next}} \in B_g$} \, $S \gets S + 1$ \EndIf
    \EndFor
\EndFor
\State \Return $S$
\end{algorithmic}
\end{algorithm}

\begin{algorithm}[h!]
\caption{Random-key calibrated detection.}
\label{alg:calibration}
\begin{algorithmic}[1]
\Require key $\mathcal{K}$, observation $\hat{b}_{1:T'}$, parameters $(w, m, n)$, calibration size $M$, level $\alpha$
\State $S_{\mathrm{true}} \gets S(\mathcal{K}, \hat{b}_{1:T'})$ \Comment{Eq.~\eqref{eq:detect-score}; via Algorithm~\ref{alg:detector-score}}
\State $\mathcal{S}_{\mathrm{null}} \gets [\,]$
\For{$r = 1, \ldots, M$}
    \State $\mathcal{K}_r \sim \mathrm{Uniform}(\mathrm{KeySpace})$
    \State $\mathcal{S}_{\mathrm{null}}.\textsc{append}\!\bigl(S(\mathcal{K}_r, \hat{b}_{1:T'})\bigr)$
\EndFor
\State $p \gets \bigl(1 + \bigl|\{S_r \in \mathcal{S}_{\mathrm{null}} : S_r \ge S_{\mathrm{true}}\}\bigr|\bigr) / (M + 1)$
\State \Return $\mathbbm{1}[p < \alpha],\; p$
\end{algorithmic}
\end{algorithm}

\section{Proofs}
\label{app:proofs}

\subsection{Proof of Theorem~\ref{thm:deletion}}

We prove the two parts separately. Throughout, let $b_{1:T}$ denote the
clean watermarked sequence and $\hat{b}$ the result of deleting $d$ positions
chosen uniformly at random, with $\rho = d/T$.

\paragraph{Proof of (a): round-indexed collapse.}
The round-indexed encoder of \citet{huang2025agentguide} biases the action at step $t$
toward the subset $G_{\mathcal{K}}(t)$, so for the clean sequence the
indicator $\mathbbm{1}[b_t \in G_{\mathcal{K}}(t)]$ has expectation
$p_1 = p_0 + \Theta(\gamma p_0(1-p_0))$ under the watermarked distribution.

After deletion, let $k^* \in \{1, \ldots, T\}$ denote the original position of
the first deleted action. For each detector position $t < k^*$, the surviving
action retains its original position index, so the round-indexed indicator
$\mathbbm{1}[\hat{b}_t \in G_{\mathcal{K}}(t)]$ continues to evaluate the
same encoder--detector pairing as in the clean sequence. Its expectation
remains $p_1$.

For each detector position $t \ge k^*$ that exists in the deleted sequence,
however, the action at detector position $t$ originated from a strictly later
encoder position $t' > t$ in the clean sequence. The encoder used
$G_{\mathcal{K}}(t')$ to bias this action, but the detector evaluates it
against $G_{\mathcal{K}}(t)$. By the pseudorandom misalignment idealization
stated in the theorem, this off-index comparison contributes only the null hit
rate:
\begin{equation}
\mathbb{E}\bigl[\mathbbm{1}[\hat{b}_t \in G_{\mathcal{K}}(t)]\bigr] \;=\; p_0,
\qquad t \ge k^*.
\label{eq:proof-collapse}
\end{equation}

Summing across positions, the expected score gap relative to a non-watermarked
sequence with expectation $p_0$ at every aligned position is
\begin{equation}
\mathbb{E}\bigl[S^{\mathrm{RI}}_{\mathrm{after}} - S^{\mathrm{RI}}_{\mathrm{null}} \,\big|\, k^*\bigr]
\;=\; (k^* - 1) \cdot (p_1 - p_0)
\;=\; (k^* - 1) \cdot \Theta(\gamma p_0(1-p_0)).
\label{eq:proof-conditional-gap}
\end{equation}

Under uniform random deletion, $k^*$ is the minimum of $d$ uniformly chosen
positions in $\{1, \ldots, T\}$. A standard computation gives
$\mathbb{E}[k^*] = (T+1)/(d+1)$, and hence
$\mathbb{E}[k^* - 1] = (T-d)/(d+1)=O(1/\rho)$ for any fixed $\rho > 0$.
Therefore,
\begin{equation}
\mathbb{E}\bigl[S^{\mathrm{RI}}_{\mathrm{after}} - S^{\mathrm{RI}}_{\mathrm{null}}\bigr]
\;=\; \frac{T-d}{d+1} \cdot \Theta(\gamma p_0(1-p_0))
\;=\; O(1/\rho).
\label{eq:proof-ag-bound}
\end{equation}

The clean-sequence gap is
$\mathbb{E}[S^{\mathrm{RI}}_{\mathrm{clean}} - S^{\mathrm{RI}}_{\mathrm{null}}]
= T \cdot \Theta(\gamma p_0(1-p_0)) = \Theta(T)$.
The ratio of post-attack gap to clean gap is therefore $O(1/(\rho T))$, which
vanishes as $T \to \infty$ for any fixed $\rho > 0$. \qed

\paragraph{Proof of (b): sliding-window deterministic bound.}
Recall the score
\begin{equation*}
S(\mathcal{K}, b_{1:T})
\;=\; \sum_{i=1}^{T-w} \sum_{j=1}^{m}
\mathbbm{1}\!\left[b_{i+w} \in G_{\mathcal{K}}(b_{i:i+w-1} \,\|\, j)\right].
\end{equation*}
Each term in the outer sum is indexed by a pair $(c, b^+)$ where $c =
b_{i:i+w-1}$ is a length-$w$ context and $b^+ = b_{i+w}$ is its immediate
successor; we refer to such a pair as a \emph{contextual transition}. The
sequence $b_{1:T}$ contains exactly $T - w$ contextual transitions, and each
contributes $m$ indicators to $S$.

Fix any deleted position $k \in \{1, \ldots, T\}$. We claim that $b_k$
participates in at most $w + 1$ contextual transitions of the clean sequence.
Indeed, $b_k$ appears in the context $b_{i:i+w-1}$ if and only if
$i \le k \le i + w - 1$, i.e., $i \in \{k - w + 1, \ldots, k\}$, giving at
most $w$ such contexts. Additionally, $b_k$ appears as the successor $b_{i+w}$
if and only if $i + w = k$, giving exactly one such transition (when
$k > w$). Summing, $b_k$ participates in at most $w + 1$ contextual
transitions and therefore contributes to at most $(w + 1)\, m$ indicators
in $S(\mathcal{K}, b_{1:T})$.

Deleting $b_k$ may invalidate at most the clean indicators in which $b_k$
participates; all other clean indicators are computed from contexts and
successors that exclude position $k$ and remain present, possibly at shifted
indices.
Each invalidated
indicator can change by at most $1$ in absolute value, and we are interested
in the worst-case decrease, which is at most $1$ per indicator. Hence
\begin{equation*}
S(\mathcal{K}, b_{1:T} \setminus \{b_k\}) \;\ge\; S(\mathcal{K}, b_{1:T}) - (w + 1)\, m.
\end{equation*}

For $d$ deletions, iterating this argument yields
\[
S(\mathcal K,\hat b)
\ge
S(\mathcal K,b_{1:T}) - d(w+1)m .
\]
This proves the deterministic additive bound. We do not convert this into a
universal multiplicative bound on the detection gap, since the clean score and
the null-calibrated score gap depend on the action distribution and the
watermark bias. \qed

\section{Experimental Setup Details}
\label{app:setup}
 
\subsection{Models and Inference}
\label{app:setup:models}
 
We use three open-weight instruction-tuned LLMs as the agent backbone, covering three families and two parameter scales:
 
\begin{table}[h]
\centering
\begin{tabular}{ll}
\toprule
Model & HuggingFace ID \\
\midrule
Llama-3.2-3B & \texttt{meta-llama/Llama-3.2-3B-Instruct} \\
Gemma-4-E4B  & \texttt{google/gemma-4-E4B-it} \\
Qwen3-4B     & \texttt{Qwen/Qwen3-4B-Instruct-2507} \\
\bottomrule
\end{tabular}
\end{table}
 
All models are loaded in bfloat16 via \texttt{transformers==5.7.0} on Python 3.10. The decoding temperature for the agent LLM is fixed at 0.7 across all experiments. We use a single agent profile, \texttt{active\_calm} (the Active Calm persona from the Agent Guide profile set), throughout to isolate watermark effects from persona-induced variance.
 
\subsection{Datasets}
\label{app:setup:datasets}
 
We evaluate on three agent benchmarks that span tool-use, embodied reasoning, and social simulation:
 
\begin{table}[h]
\centering
\resizebox{\textwidth}{!}{%
\begin{tabular}{lll}
\toprule
Dataset & Scale & Episode setting \\
\midrule
ToolBench & 200 queries (50 for $\gamma$-sweep) & ReAct loop, \texttt{max\_steps\_per\_episode}\,$=$\,6 \\
ALFWorld & 50 tasks & \texttt{max\_episode\_length}\,$=$\,50, \texttt{num\_few\_shot}\,$=$\,1 \\
OASIS Reddit & 100 simulation steps & 1 user agent + 3 NPC agents, persona \texttt{tech\_geek} \\
\bottomrule
\end{tabular}%
}
\end{table}
 
All three benchmarks share the same agent profile (\texttt{active\_calm}) and the same watermark plumbing. Per-step behavior sets are dataset-specific: ToolBench surfaces the candidate tool list at each ReAct step, ALFWorld exposes the admissible-action list from the textual environment, and OASIS Reddit uses a fixed action vocabulary of social actions (e.g., create post, like, comment, follow).
 
\subsection{Watermark Hyperparameters}
\label{app:setup:hparams}
 
Unless otherwise stated, all experiments use the following configuration:
 
\begin{table}[h]
\centering
\begin{tabular}{lll}
\toprule
Symbol & Value & Meaning \\
\midrule
$w$ (\texttt{window\_size}) & 3 & History window for context-conditioned seeding \\
$n$ (\texttt{n\_guide}) & 3 (OASIS: 2) & Size of each guided subset $B_g^{(j)}$ \\
\texttt{n\_min} & 2 (OASIS: 1) & Minimum subset size after clipping \\
$m$ (\texttt{m\_channels}) & 8 & Number of redundant channels \\
$\gamma$ (\texttt{gamma\_mc}) & 2.0 & Bias strength in $P^{wm} \propto \max(P, \delta) \cdot \exp(\gamma \cdot s/m)$ \\
$\delta$ (\texttt{delta}) & 0.2 & Floor applied to $P$ before the exponential bias \\
$M$ (\texttt{M\_calibration}) & 1000 & Random-key calibration trials per detection \\
$\alpha$ (\texttt{detection\_alpha}) & 0.01 & $p$-value threshold for the headline detection result \\
z-threshold & 2.0 & Threshold for the reference single-channel z-test \\
\bottomrule
\end{tabular}
\end{table}
 
OASIS Reddit uses smaller $n$ and \texttt{n\_min} because its action vocabulary is narrower than ToolBench's per-step tool list or ALFWorld's per-step admissible-action set. All other parameters are shared across datasets to keep the comparison clean.
 
\subsection{\texorpdfstring{$\gamma$}{gamma} Sensitivity Sweep}
\label{app:setup:gamma}
 
For the $\gamma$-sensitivity experiment we sweep $\gamma \in \{0.5, 1.0, 1.5, 2.0, 2.5, 3.0\}$ with $m_{\text{channels}} = 8$ fixed. Each $\gamma$ value runs on a 50-query ToolBench subset (\texttt{configs/sweep\_react.json}), generating both the no-watermark baseline trajectory and the multi-channel watermarked trajectory from the same query set per worker.
 
\subsection{Computational Resources}
\label{app:setup:compute}
 
All experiments run on a SLURM-managed cluster with two relevant GPU partitions:
 
\begin{itemize}
    \item \texttt{suma\_a100}: $2 \times$ NVIDIA A100 GPUs, used for the main rollout experiments (main detection table, ablations, $\gamma$-sweep, cross-model runs).
    \item \texttt{suma\_pro6000}: $2 \times$ NVIDIA RTX PRO 6000 Blackwell GPUs, used for sanity checks and post-hoc analysis sweeps (such as the deletion-attack re-detection, which does not require LLM inference).
\end{itemize}
 
A single full main run (200 ToolBench queries, all modes side by side under one rollout) completes in roughly 4 to 6 hours of A100 wall time per agent LLM.


\newpage

\end{document}